\documentclass[aps,pre,amssymb,twocolumn,amsmath,superscriptaddress,showpacs,10pt]{revtex4}

\usepackage{hyperref}
\usepackage{graphicx}
\usepackage{dcolumn}
\usepackage{bm}
\usepackage{color}
\usepackage{dsfont}

\begin{document}

\newcommand{\beq}{\begin{equation}}
\newcommand{\eeq}{\end{equation}}
\newcommand{\barr}{\begin{eqnarray}}
\newcommand{\earr}{\end{eqnarray}}

\def\bra#1{\langle{#1}|}
\def\ket#1{|{#1}\rangle}
\def\sinc{\mathop{\text{sinc}}\nolimits}
\def\cV{\mathcal{V}}
\def\cH{\mathcal{H}}
\def\cT{\mathcal{T}}
\def\cS{{\cal S}}
\renewcommand{\Re}{\mathop{\text{Re}}\nolimits}
\newcommand{\tr}{\mathop{\text{Tr}}\nolimits}

\definecolor{dgreen}{rgb}{0,0.5,0}
\newcommand{\green}{\color{dgreen}}
\newcommand{\RED}[1]{{\color{red}#1}}
\newcommand{\BLUE}[1]{{\color{blue}[[per Angelo: #1]]}}
\newcommand{\GREEN}[1]{{\color{dgreen}#1}}
\newcommand{\REV}[1]{{\color{red}[[#1]]}}
\newcommand{\KY}[1]{\textbf{\color{red}[[#1]]}}
\newcommand{\SP}[1]{{\color{blue} [[#1. Saverio]]}}
\newcommand{\rev}[1]{{\color{red}[[#1]]}}

\def\HN#1{{\color{magenta}#1}}
\def\DEL#1{{\color{yellow}#1}}

\title{Language discrimination and clustering via a neural network approach}

\author{Angelo Mariano}
\affiliation{ENEA, Italian National Agency for New Technologies, Energy and Sustainable Economic Development, Strada Statale 7 Appia km 706, I-72100 Brindisi, Italy} 

\author{Giorgio Parisi}
\affiliation{Dipartimento di Fisica, Universit\`{a} di Roma ``Sapienza'', Piazzale Aldo Moro 2, I-00185 Roma, Italy}
\affiliation{Centre for Statistical Mechanics and Complexity (SMC), CNR-INFM, I-00185 Roma, Italy}
\affiliation{INFN, Sezione di Roma,  I-00185 Roma, Italy}

\author{Saverio Pascazio}
\affiliation{Dipartimento di Fisica and MECENAS, Universit\`a di
Bari, I-70126 Bari, Italy} \affiliation{INFN, Sezione di Bari,
I-70126 Bari, Italy}

\begin{abstract} 
We classify twenty-one Indo-European languages starting from written text. We use neural networks in order to define a distance among different languages, construct a dendrogram and analyze the ultrametric structure that emerges. Four or five subgroups of languages are identified, according to the ``cut" of the dendrogram, drawn with an entropic criterion. The results and the method are discussed.
\end{abstract}


\maketitle

\section{Introduction and Motivations}

The identification and classification of languages is a complex and interesting subject, that has roots in many areas of sciences and humanities. This problem can be tackled in different ways. We try here an approach that hinges on 
written text and makes use of neural networks. There is a vast literature on this and related subjects, ranging from artificial intelligence and computer science \cite{compscience}, to linguistics \cite{linguistic} and the physical literature \cite{BCL}. We shall hinge on physical and mathematical concepts and ideas.

\begin{figure}
\centering
\includegraphics[width=8 cm]{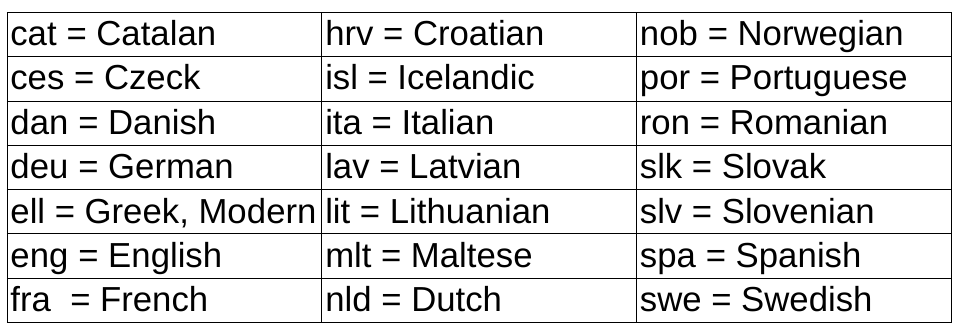}\\
\caption{Languages and acronyms.}
\label{acronyms}
\end{figure}

Our main objective is to define a distance among a given set of languages and identify sub-groups of languages that are more similar. In our case study, the set is made up of the 21 Indo-European languages listed in Fig.\ \ref{acronyms}. The strategy consists in asking a feed-forward neutral network to distinguish them two by two. The error will be taken as a measure of the similarity between the two languages: a large error signifies closeness between the two languages, while a small error indicates a large separation.

The main idea is to simulate the learning stage of an unexperienced speaker (or even a child). Our starting point is the assumption that a beginner who is learning (say) English will find German more familiar (closer) than Portuguese. If asked to discriminate English from German she/he/it will make more mistakes than when discriminating English from Portuguese. We shall mimic the above-mentioned learning stage through an artificial neural network.
The use of the latter in cognitive sciences \cite{neuralnet} and in fields of investigation that can be considered  ``complex" is often inter-disciplinary. Among related areas of study there is artificial intelligence inspired by biology \cite{bioinspired}, DNA sequence classification and disease classification and diagnoses \cite{BP}.

Our focus will be on physical ideas and concepts, and in particular on how physical notions, such as that of entropy, enable one to identify and classify sub-groups of languages that share some similarity. Since our method hinges on written text, the rules of alphabet transcription will play an important role and (presumably) have an influence on the final classification \cite{KT}. This motivated us to look at a group of 21 Indo-European languages written in Roman characters, with two exception: Greek and Maltese. Maltese is written Roman characters, but unlike the other languages in the group, is a Semitic language. Greek is Indo-European, but is written in a different script. It will be interesting to see which of these aspects prevails and how these two languages are classified.

This article is organized as follows.
In Sec.\ \ref{sec:NN} we introduce the method of investigation, based on neural networks. Languages are discriminated with some error, and this enables one to define a distance. The features of this error are analyzed in Secs.\ \ref{sec:error} and \ref{sec:shuffle}, where the sentences are modified and manipulated in order to test the method of discrimination. In Sec.\ \ref{sec:linkage} we recall the mathematical definition of distance and describe the linkage algorithm, introducing the cophenetic coefficient. We proceed to classify languages in Sec.\ \ref{sec:dist_lang} and discuss the tree and ultrametric structure that emerge in Sec.\ \ref{sec:tree_ultra}. We conclude in Sec. \ref{sec:concl}.
The methods utilized, the learning phase and the code are detailed in Appendix \ref{sec:code}.

\section{Neural Network}
\label{sec:NN}

\subsection{Method}
\label{sec:method}
We use a feed-forward neural network with back-propagation \cite{backprop}. The objective of each ``run" is to discriminate two given languages of the set. The neural network is fed with 20,000 sentences (10,000 per language), taken from the Leipzig Corpora Collection \cite{Leipzig} and different in length.

The first 40 characters are extracted from each sentence; spaces, accents, numbers and punctuation are removed and special (non ASCII) characters are replaced by their ASCII counterpart (e.g.\ French and Spanish ``\c c" is replaced by ``c" and German ``\ss " by ``ss"). These replacements make languages more similar and language discrimination more difficult, so that the task of the neural network is harder. 

Each character is encoded in a 26-dimensional vector \cite{jensen,langident}. Other types of input encoding offer no guarantee on the correct association of the weights of the synapsis to different characters and the sensibleness of the ensuing language discrimination. We avoided the use of features extracted from sentences, that make use of \emph{a priori} assumptions and can introduce some bias. (We are interested in classifying data when no a priori knowledge is given or available, as in some phylogenetic analyses of biological sequences \cite{cavallisforza}.)

\begin{figure}
\centering
\includegraphics[width=8 cm]{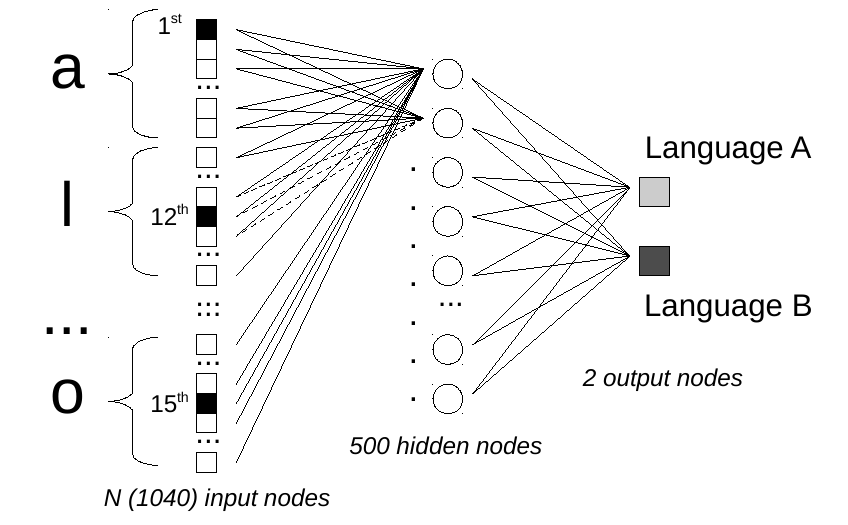}\\
\caption{The discrimination scheme. A 40-character-long sentence is fed to the neural network, whose input, central and output layers consist of $1040 (=26 \times 40)$, 500 and 2 nodes respectively.
After a training period, during which 16,000 sentences in two languages are fed to the network, the two output nodes yield the probability that a given sentence is classified as belonging to one of the two languages. Spaces, accents and punctuation are removed, and the sentence is always cut at 40 characters, so that ``Alice thought she might as well wait, as she had nothing else to do, and perhaps after all it might tell her something worth hearing.", becomes ``alicethoughtshemightaswellwaitasshehadno".
}
\label{alice}
\end{figure}

The neural network is made up of three layers. The first layer contains $1040 (=26 \times 40)$ input nodes, the central layer 500 nodes, the output layer 2 nodes. Each time, one language is discriminated against another language: the two output nodes give the probability of correctly interpreting the input. The scheme is depicted in Fig.\ \ref{alice}. 

The results are averaged according to a 5-fold cross-validation technique. Specifically, our validation set is made up of 4,000 sentences. The neural network makes an error when, if fed with language $A$, it mistakenly answers $B$ (or \emph{vice versa}). A training period, that minimizes an appropriate cost function, is repeated against the remaining 16,000 sentences, until the error against the validation set is steady. In this way, a set of 5 errors is obtained. The labels $A$ and $B$ are then interchanged and after a new training period, 5 additional errors are otained. The error $\epsilon$ used in the following is the average over the afore-mentioned 10-fold run.
The training and code utilized are detailed in Appendix \ref{sec:code}. 

\subsection{Praising errors}

The afore-mentioned neural network works very efficiently, making language discrimination very effective. After the training period, the error is always small (typically a few percent, becoming at most $20\%$ when the two languages are very similar). 

We shall make use of the error $\epsilon$ to define a distance $d$ between the two languages, according to the formula
\begin{equation}
d = \frac{1}{\epsilon} -1 \in \mathbb{R}_+ .
\label{distanceerror}
\end{equation}
The above definition entails a certain degree of arbitrariness, as any monotonic function of $1/\epsilon$ would be equally suited. One advantage is that $d$ takes values on the positive real half-line $\mathbb{R}_+=[0,\infty)$.

At each run, the neural network discriminates two given languages in the set. When the error and the distance between these two languages are obtained, a new run starts: the network is reset and a new couple of languages is used. This procedure yields a matrix of distances. Before embarking in the description of this task, we shall test the system and try to elucidate its functioning and characteristics.

\section{Error vs input length}
\label{sec:error}

\begin{figure}
\centering
\includegraphics[width=8 cm]{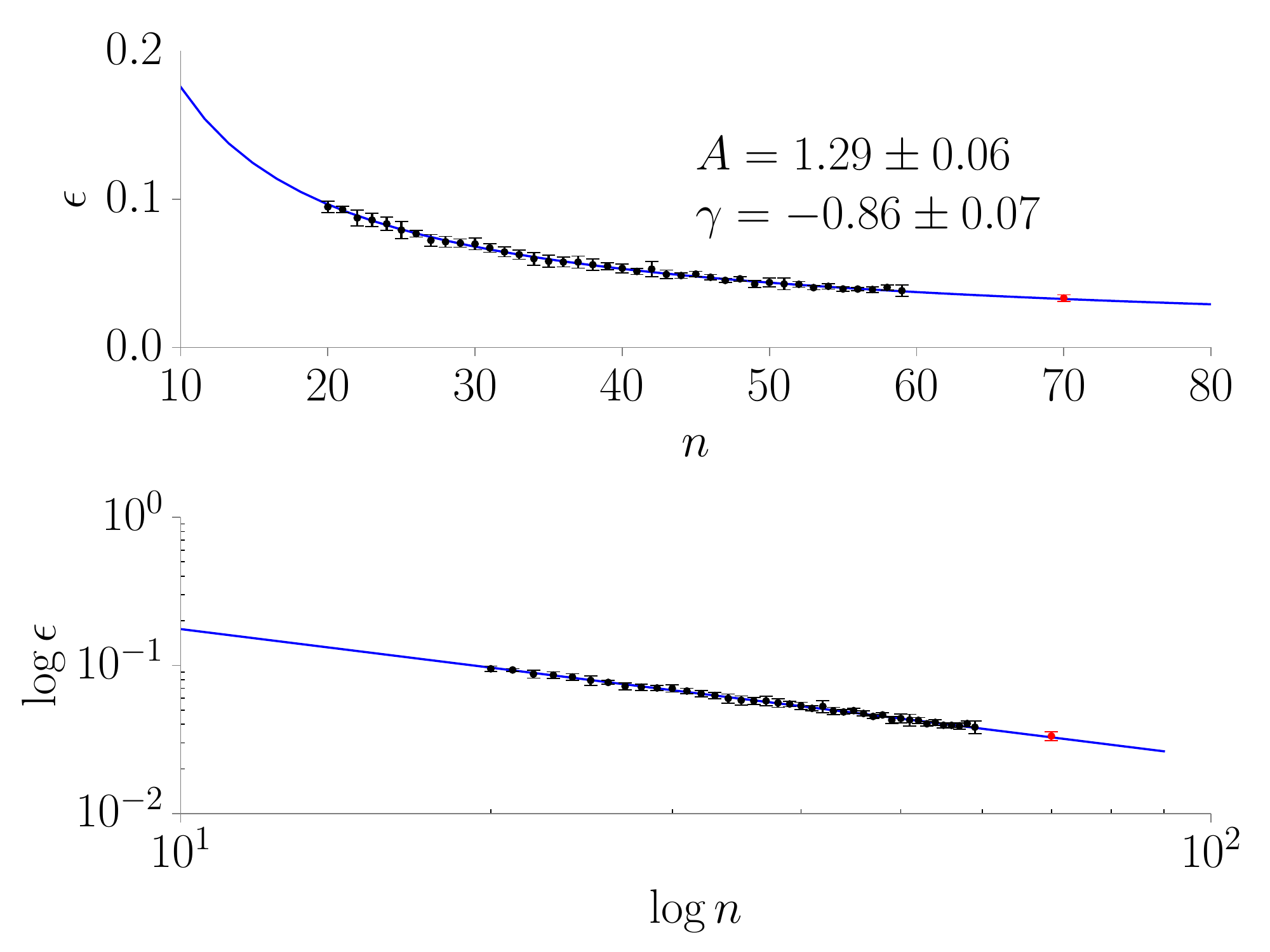}\\
\caption{English vs Italian: error $\epsilon$ (percent) vs length $n$ of input sentence. Upper panel: linear plot. Lower panel: logarithmic plot. The error bars are the standard deviations over the cross-validation ensemble of the neural network. (The point at $n=70$ is not taken into account for the fit.)}
\label{errorvsinputNpoint}
\end{figure}

In the preceding section and throughout this article, the length $n$ of the input sentences is taken to be $n=40$. This is dictated by efficiency and by the duration of the computer simulation.
In Figure \ref{errorvsinputNpoint} the error is plotted as a function of the length of the input sentence $n$, for $n=30, \ldots, 60$. We are discriminating here English vs Italian. 

The first layer of the neural network changes, because the number of input nodes $N$ is taken to be
\begin{equation}
N = 26 \times n,
\end{equation}
where $n$ is the length of input sentence. The middle and output layers do \emph{not} change.
The fit yields a power law
\begin{equation}
\epsilon = A n^\gamma,
\label{errorvsX}
\end{equation}
with $A = 128.59 \pm 6.02$ (percent), $\gamma=-0.86 \pm 0.07$.
This enables one to assert that if one has enough computational power, the error in discriminating two languages can be made arbitrarily small. 
In order to corroborate the analysis, \emph{after} determining the parameters $A$ and $\gamma$, we added a point at $n=70$. This additional point is very accurately fitted, validating the ansatz (\ref{errorvsX}).

\begin{figure}
\centering
\includegraphics[width=8 cm]{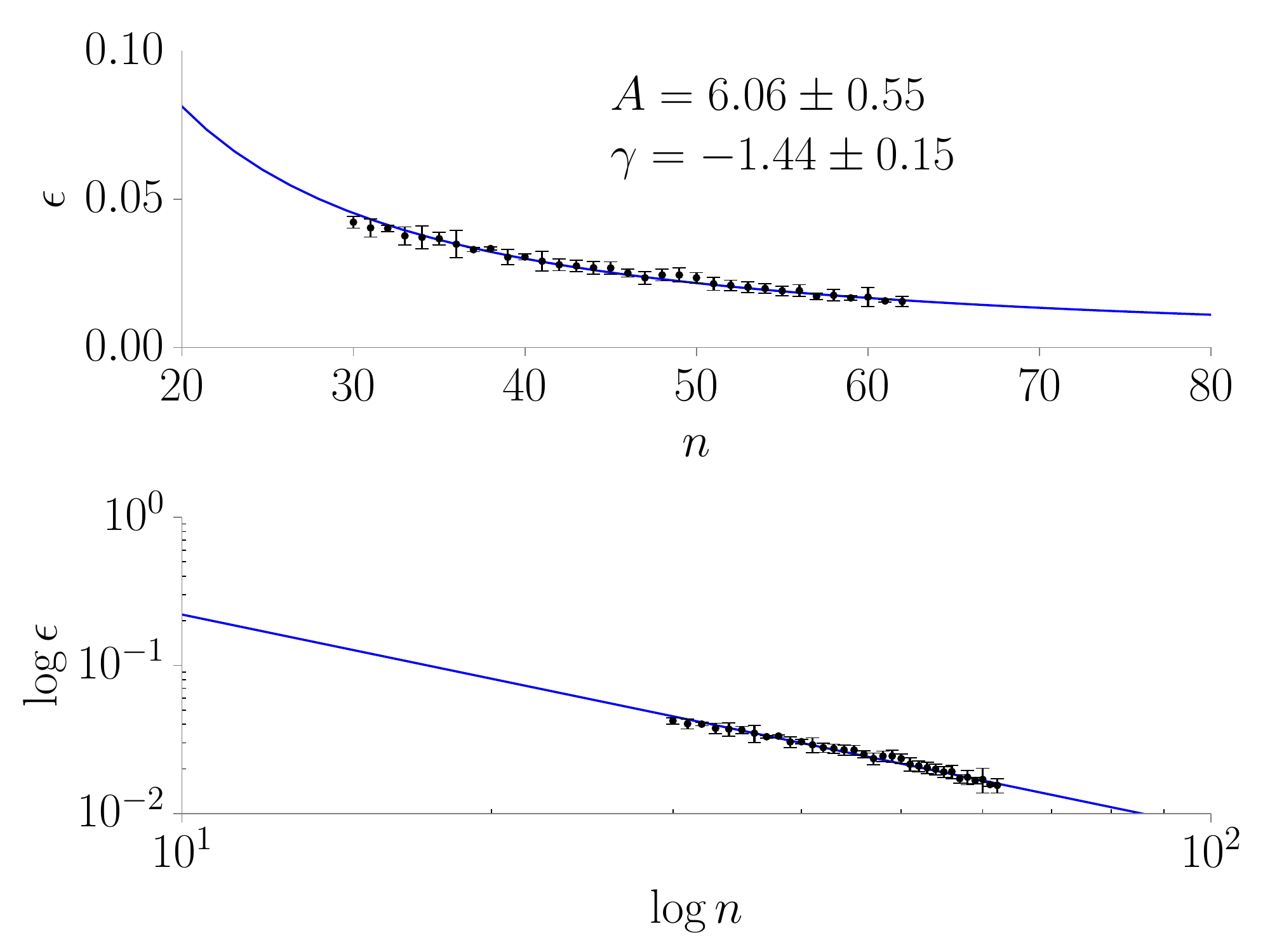}\\
\caption{French vs Swedish: error $\epsilon$ (percent) vs length $n$ of input sentence. Upper panel: linear plot. Lower panel: logarithmic plot. The error bars are the standard deviations over the cross-validation ensemble of the neural network.}
\label{errorvsinputNfraswe}
\end{figure}

One may wonder whether the coefficients of the fit ($\gamma$ in particular) are language-dependent. 
In Figure \ref{errorvsinputNfraswe} the error is plotted as a function of the length of the input sentence $n$ when the network discriminates French vs Swedish. The fit always yields a power law, but the coefficients are different.

\section{Scrambling the characters}
\label{sec:shuffle}

One could conjecture that the neural network is simply counting the frequency of the letters in a given language. In order to test this hypothesis, we randomly ``scrambled" the letters of the 20,000 input sentences (10,000 $\times$ two languages) and asked the network to discriminate the two languages. 

\begin{figure}[h]
\centering
\includegraphics[width=8 cm]{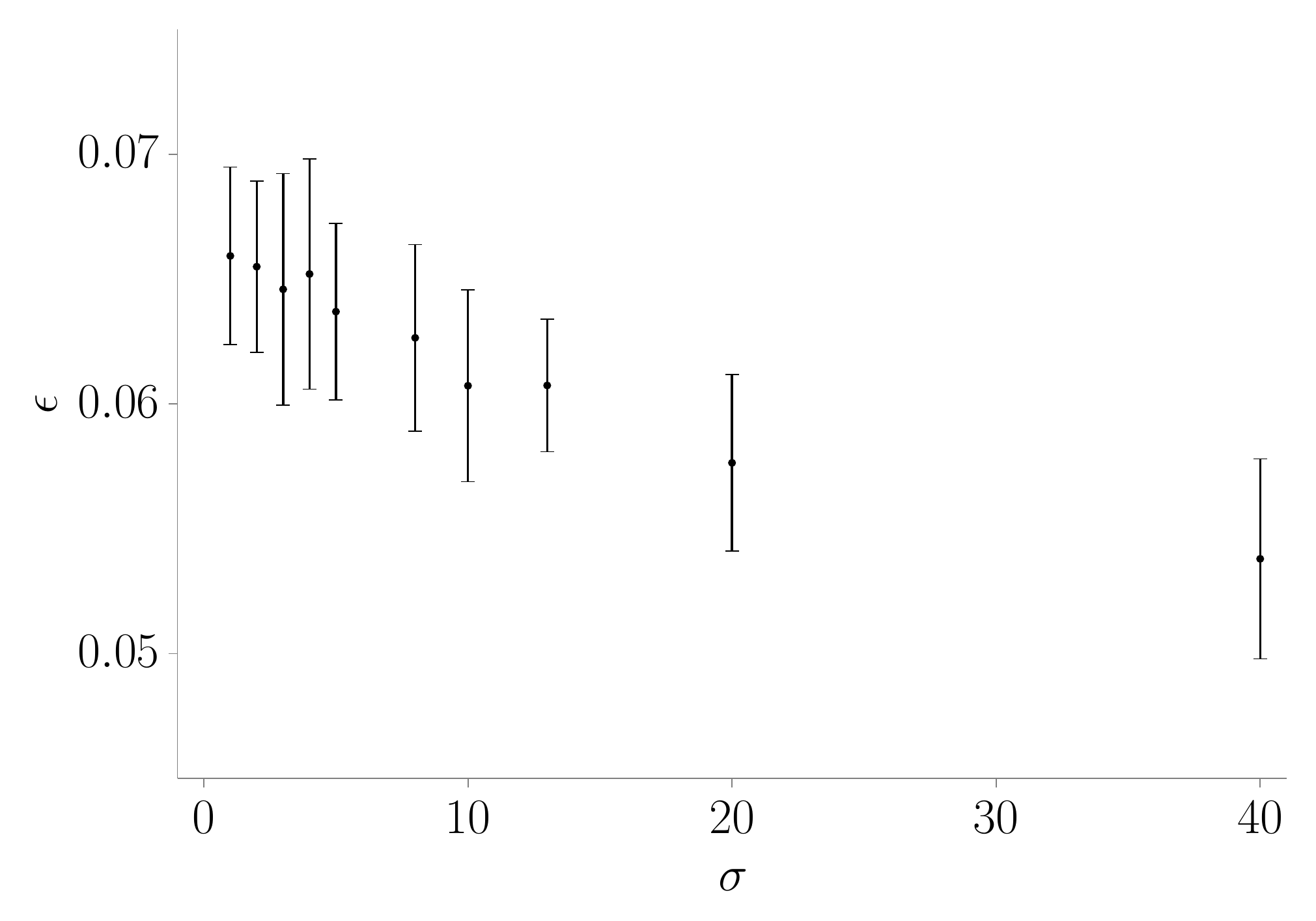}\\
\caption{Errors $\epsilon$ vs size $\sigma$ of the scrambled ``cell". The error bars are the standard deviation over 8 repetitions of the cross-validation ensemble of the neural network. (Italian vs English)}
\label{shuffle}
\end{figure}

The result for English vs Italian is displayed in Fig.\ \ref{shuffle}: $\sigma = 1$ means that the sentence is randomly scrambled character by character, so that the neural network in this case simply ``counts" the frequency of characters in each language;  
$\sigma = 2$ means that the sentence is scrambled in ``cells" made up of couple of characters, and so on; $\sigma = 40$ means that the sentence is not scrambled. 

The points plotted in Fig.\ \ref{shuffle} are (close to) the divisors of 40 (length of the input sentences). When the division is not exact, a number of zeros is added to the last character and removed at the end of the procedure. Example: for the point with abscissa = 3 we added two 0's at the end of the (40-character-long) sentence, so that $40 + 2 = 42$ becomes divisible by 3, and there are 14 triples of characters; the two spurious zeros are removed at the end of the procedure. 

\begin{figure}
\centering
\includegraphics[width=8 cm]{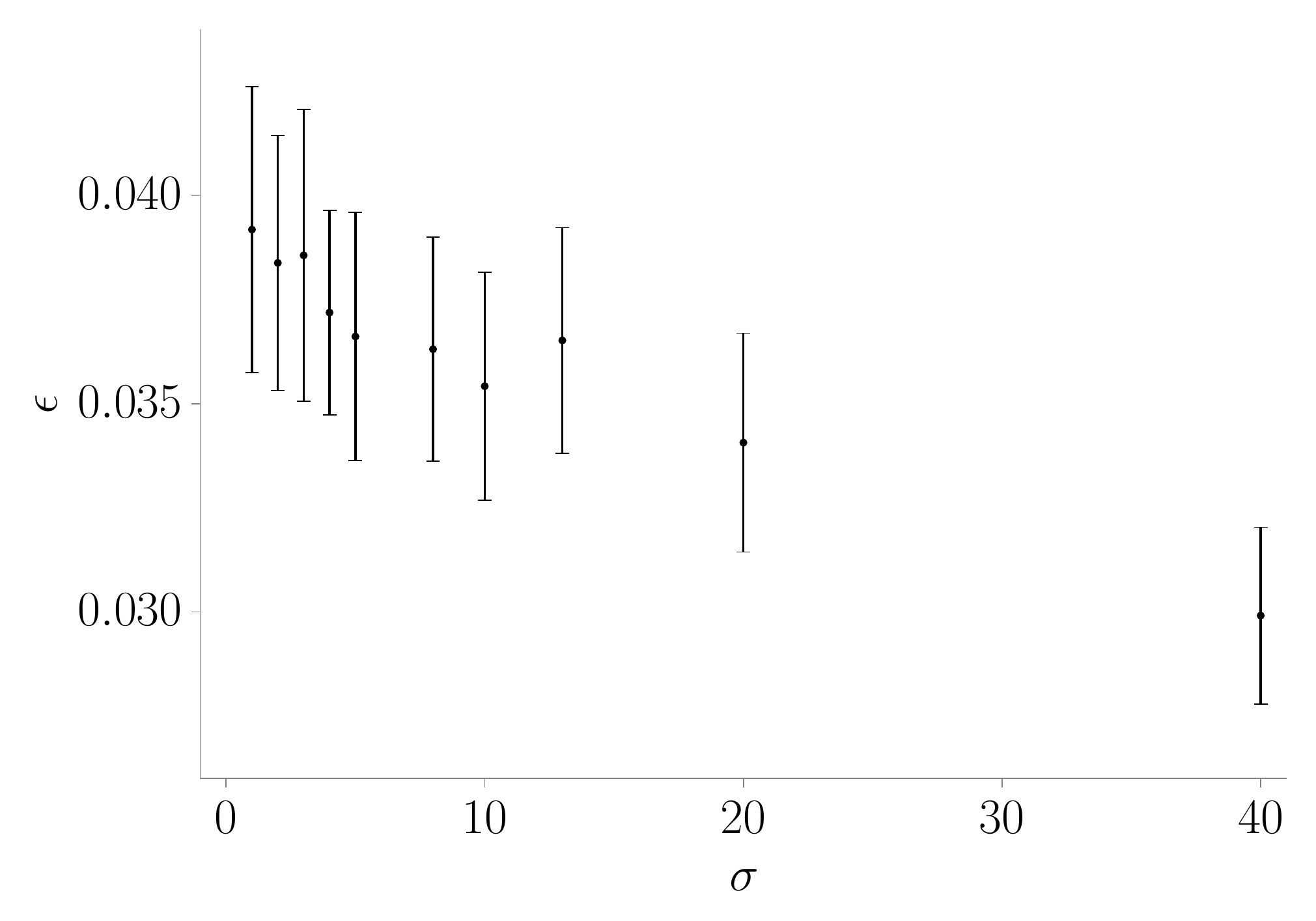}
\caption{Errors $\epsilon$ vs size $\sigma$ of the scrambled ``cell". The error bars are the standard deviation over the cross-validation ensemble of the neural network. (French vs Swedish)}
\label{shufflefraswe}
\end{figure}

The whole procedure aims at understanding whether the network is able to detect the presence of ``correlations" with longer range within a language and whether it makes use of this information when performing its discrimination task.
It is clear that the error monotonically decreases with $\sigma$: the network detects longer-range correlations. At the same time, the statistics is not sufficient to draw clear-cut conclusions about the $\sigma$-dependence of $\epsilon$.
For this reason, in Fig.\ \ref{shufflefraswe} the same procedure was repeated when discriminating French vs Swedish. The trend is analogous. 

\section{Distance, linkage and dendrograms}
\label{sec:linkage}

We now briefly recall some mathematical notions. We start by recalling the definition of distance, then describe the linkage algorithm, finally introduce the cophenetic coefficient. We proceed to classify languages in the next section.

\subsection{Distance}
\label{sec:distance}

A metric $d$ on a set ${\cal S}$ of points is a non-negative application (distance)
\beq 
d: {\cal S} \times {\cal S}
\longrightarrow \mathbb{R}_+ , \label{distfunc}
\eeq
where
$\mathbb{R}_+=[0,\infty)$, endowed with the following properties,
valid for each couple of points $x,y \in \cS$:
\barr
& & d(x,y)=0 \quad \Longleftrightarrow \quad x=y, \label{assio_1} \\
& & d(x,y)=d(y,x), \label{symmetry} \\
& & d(x,y)\leq d(x,z)+d(z,y) . \quad \forall z \in{\cal S} \label{triangular}
\earr
A metric is called an ultrametric if it satisfies the following stronger version of the triangle inequality (\ref{triangular})
\beq 
d(x, z) \leq \max(d(x, y), d(y, z)) .
\label{ultram}
\eeq
Ultrametrics are very useful when one endeavors to unearth the structure of a (phylogenetic) tree, 
yielding a taxonomic classification of the points in the set ${\cal S}$ \cite{MPV}.

It is far from being obvious that definition (\ref{distanceerror}) satisfies the axioms (\ref{assio_1})-(\ref{triangular}) of a \emph{bona fide} mathematical distance. More so for property (\ref{ultram}).

\subsection{Linkage algorithms}
\label{sec:Linkage}

In order to classify the 21 languages listed in Fig.\ \ref{acronyms} and identify subgoups in the set, we need a linkage algorithm. Linkage algorithms yield a cluster structure that is displayed in the form of a tree or dendrogram \cite{jain}. 
We will adopt an agglomerative algorithm, by linking the clusters through an iterative process. The original data set ${\cal S}$ is made up of $m=21$ elements (points). At the first level (leaves of the dendrogram) the number of classes in the data set is $m= 21$.  At the first iteration the two closest elements are clustered together, reducing the number of classes to $m-1=20$ (if more than two elements are at the closest distance, we pick a random couple among them). At the second iteration one defines a new distance $d'$ between the remaining elements of ${\cal S}$ and the first cluster formed. The distances are then recalculated and the two closest objects are joined. 
At the following iterations one must define a new distance among points and clusters. After $m$ steps, all the points are grouped together in one single cluster, corresponding to the whole data set. 

A number of linkage algorithms can be adopted, that differ in the definition of ``distance" between subsets of points. 
Such a ``distance" is usually not a distance according to the mathematical definition given in Sec.\ \ref{sec:distance}, but rather a ``dissimilarity" criterion \cite{BBDFPP}. This criterion can be given in a variety of different ways and entails some elements of arbitrariness. 
Some linkage algorithms make use of the coordinates of the points in the set, while some others rely only on distances, and are to be preferred in our case.

Among the most common methods there are the ``single", ``complete", ``average", Ward and Hausdorff linkage \cite{cox,fisher,ward,BBDFPP}. In our case-study (see next section) all criteria yield similar results, although the single linkage suffers from the well-known chaining effect. We will display only the results obtained with the average linkage, as it behaves better in terms of the cophenetic coefficient, to be introduced in the following subsection.
 
The average linkage is based on the following definition of ``distance" between two clusters $U$ and $V$
\beq 
d(U,V)=\sum_{\substack{i \in U \\ j \in V}} \frac{d(i,j)}{|U| * |V|}, 
\label{average}
\eeq 
where the summations run over the elements of the clusters and $|U|$ and $|V|$ represent the cardinality of cluster $U$ and $V$. As mentioned before, (\ref{average}) is not a distance in the mathematical sense, but should rather be viewed as a ``dissimilarity" index \cite{BBDFPP}.

\begin{figure*}
\centering
\includegraphics[width=\textwidth]{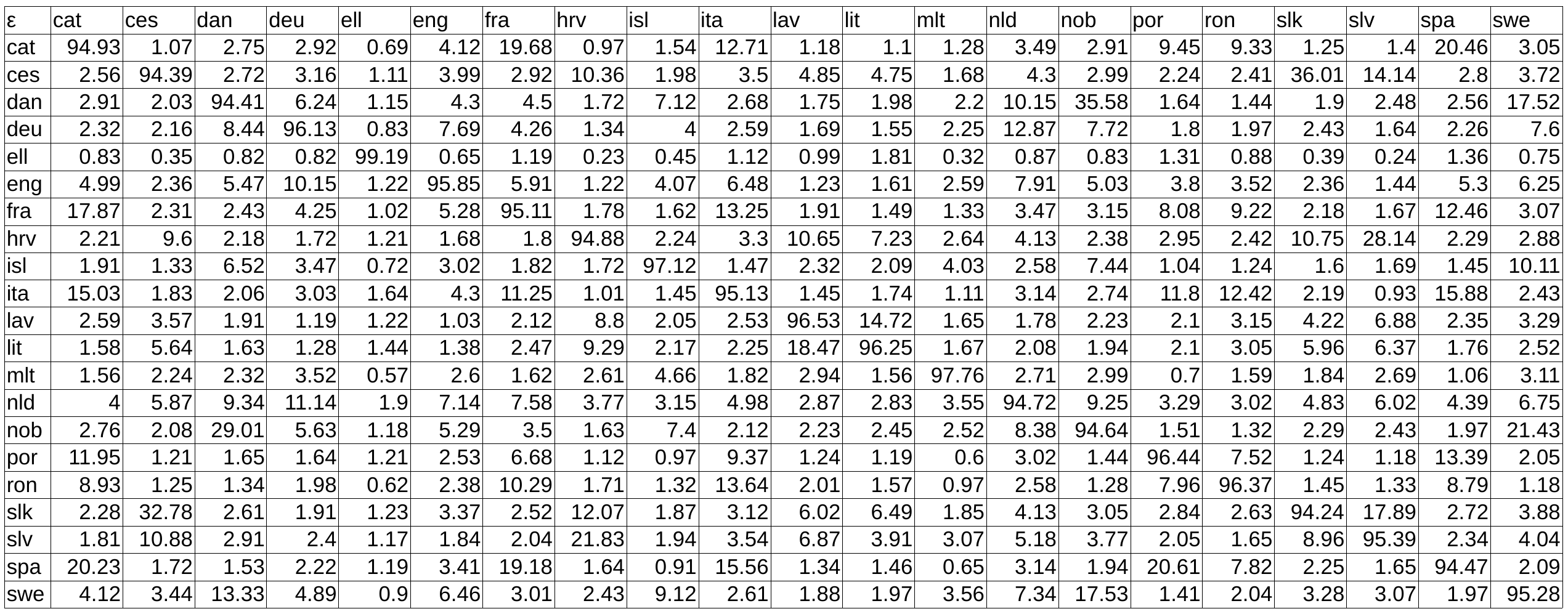}\\
\includegraphics[width=\textwidth]{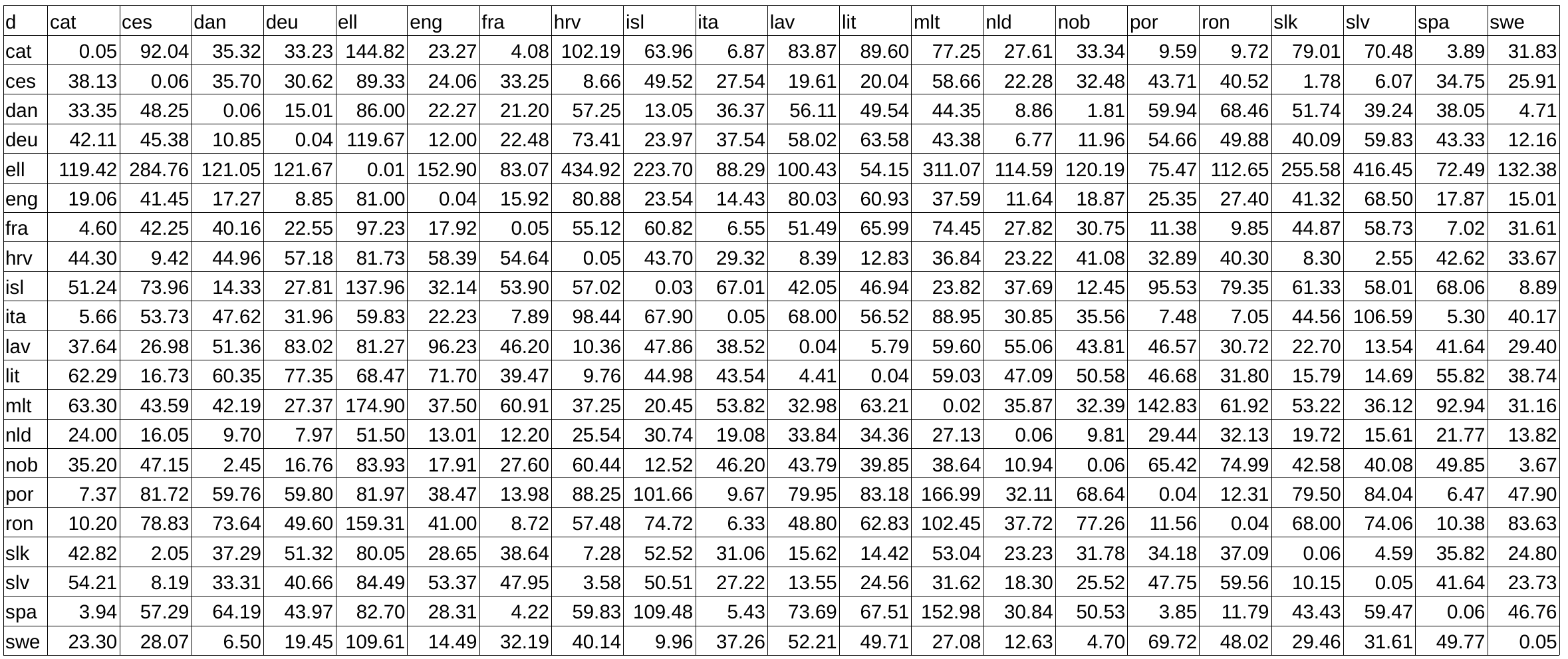}\\
\caption{Error $\epsilon$ and distance $d$, according to Eq.\ (\ref{distanceerror}). 
The input language is in the first column, while the output language is in the first row. So for example the entry $\epsilon$(dan,isl)= 7.12 in the upper table is obtained by feeding (about) 2,000 Danish (dan) sentences to the network that (mistakingly) recognizes them as Icelandic (isl) with probability $7.12\%$ (average over 10 runs, see Sec.\ \ref{sec:method}). By contrast, the entry $\epsilon$(isl,dan)=6.52 is obtained by feeding (about) 2,000 Icelandic sentences to the network that (mistakingly) recognizes them as Danish with probability $6.52\%$.
The diagonal in the upper panel gives the mean percentage of correct identifications. }
\label{table1}
\end{figure*}

\subsection{Cophenetic coefficient}
\label{sec:cophenetic}

The cophenetic coefficient \cite{compare_dendro} is a measure of how faithfully a dendrogram preserves the pairwise distances between the points of the original data set. It enables one to judge how close the original metric is to an ultrametric. It is defined as a correlation coefficient between original distance $d$ and cophenetic distance $d_c$ on the dendrogram, defined as the distance at which two leaf nodes $i,j$ are clustered together. 

The cophenetic coefficient $c$ is defined as
\beq
c=\frac{\sum_{i<j} (d(i,j)-\langle{d}\rangle)(d_c(i,j)-\langle{d_c}\rangle)}
{\sqrt{\sum_{i<j} (d(i,j) -\langle{d}\rangle)^2 \sum_{i<j} (d_c(i,j) -\langle{d_c}\rangle)^2}},
\label{cophenet}
\eeq
where $\langle d \rangle$ and $\langle d_c \rangle$ are the mean original distance and the mean cophenetic distance, respectively, and the sum is over all the nodes of the dendrogram.
A value of $c$ close to 1 signifies the presence of a good ultrametric structure.

\section{Distances among languages}
\label{sec:dist_lang}

We are now ready to discriminate and classify languages, by utilizing the method described in the preceding sections. Errors $\epsilon$ and distances $d$, defined according to Eq.\ (\ref{distanceerror}), are displayed in Fig.\ \ref{table1} for the set of languages shown in Fig.\ \ref{acronyms}. A few comments are in order.

Error and distance are not symmetric, so that property (\ref{symmetry}) is not always verified. This effect becomes more important when the error is close to zero, as this induces large distances, according to Eq.\ (\ref{distanceerror}). This will affect the distances of Greek and Maltese in particular, as these two languages are the most ``different" within the group. Henceforth, the distance will be re-defined according to the symmetrized formula
\begin{equation}
d_s = \frac{2}{\epsilon_1 + \epsilon_2} -1 ,
\label{avdistanceerror}
\end{equation}
where $\epsilon_{1,2}$ are the two entries (for each couple of languages) in the upper panel of Fig.\ \ref{table1}. 

\begin{figure}
\centering
\includegraphics[width=9cm]{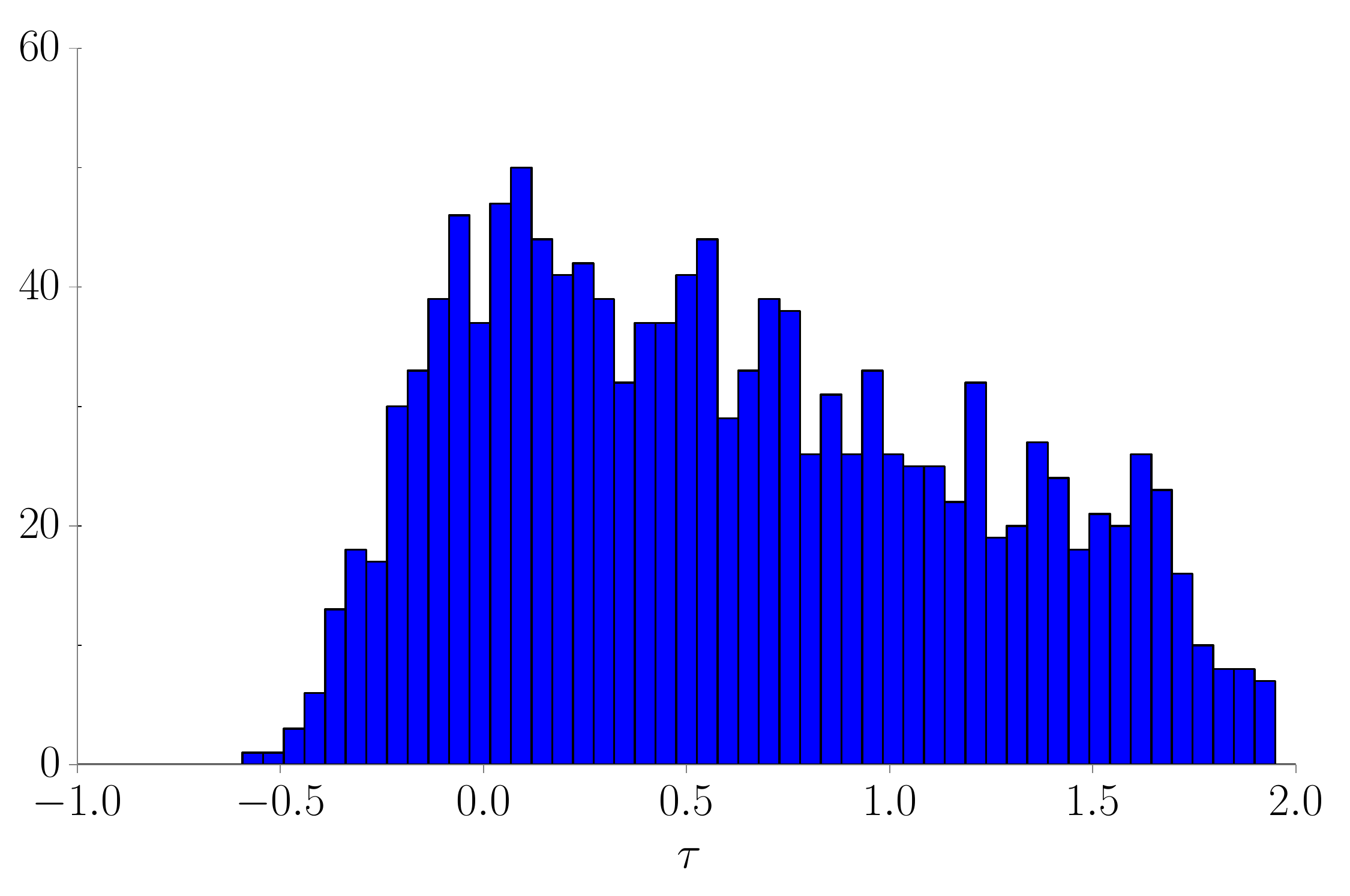}\\
\caption{Distribution function of the quantity (\ref{index}) for 1330 triads of distances taken from Fig.\ \ref{table1}.
The parameter (\ref{index}) takes values in the range $[-1, 2]$ and becomes negative if the triangle inequality (\ref{triangular}) is violated. }
\label{viola_triangol}
\end{figure}

Let us now check the validity of the triangle inequality. The histogram in Fig.\ \ref{viola_triangol} displays the frequency distribution of the quantity
\beq
\tau = \frac{d(A,C) + d(C,B) - d(A,B)}{d_\textrm{max}} ,
\label{index}
\eeq
where $A, B$ and $C$ are languages in the set and the denominator is the max of the quantities appearing in the numerator. The parameter $\tau$ takes values in the range $[-1, 2]$. It is positive for a \textit{bona fide} mathematical distance and becomes negative if the triangle inequality (\ref{triangular}) is violated. 
The triangle inequality is not satisfied in about 17\% of the cases. Its violation is usually small (less than $-0.27$ with probability 5.2\%). These figures do not change much if one takes the unsymmetrized definition (\ref{distanceerror}). We consider this a very satisfactory feature of the method we propose.

The diagonal in the upper panel of Fig.\ \ref{table1} displays the mean percentage of correct identifications (the probability of correctly identifying the input language). Unlike the off-diagonal entries, this average is evaluated on the \emph{whole} dataset and therefore depends on the entire set of languages. A language that has many similar languages (within the given set) will have a lower mean percentage of correct identifications, while a language that 
is very different from the other languages in the given set will have a larger mean percentage of correct identifications. This will induce a ``size of the point", according to (\ref{distanceerror}) or (\ref{avdistanceerror}), in violation of property (\ref{assio_1}). Rather than considering this a negative aspect, we shall view this ``size" as a measure of the ``overlap" among the given language and other languages in the set. (This size is reproduced in Fig.\ \ref{geomap} in the following.)

\begin{figure*}
\centering
\includegraphics[width=12cm]{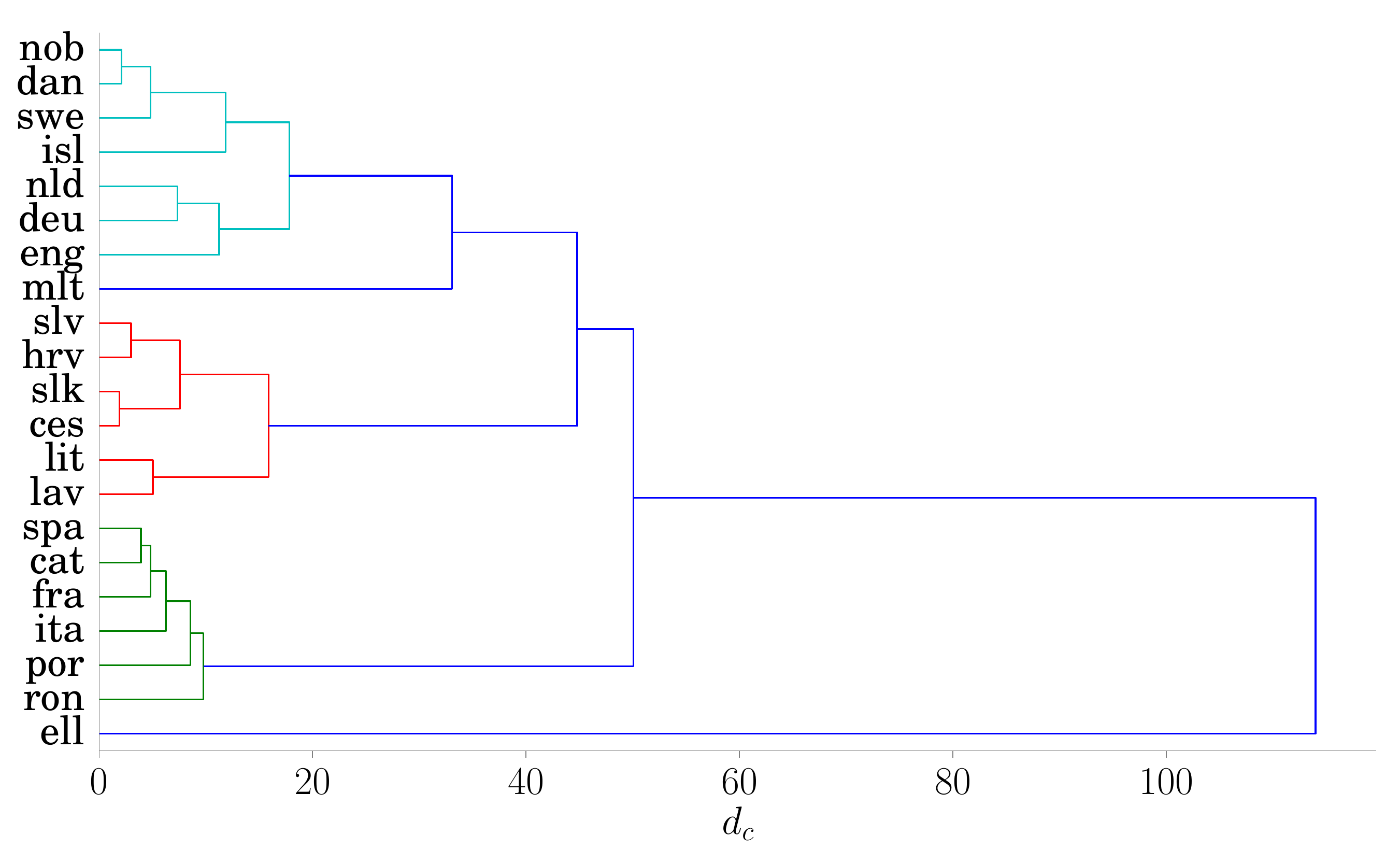}\\
\caption{The language tree}
\label{linkage_alg}
\end{figure*}

\section{Tree and ultrametric structure}
\label{sec:tree_ultra}

Figure \ref{linkage_alg} displays the dendrogram obtained by starting from the symmetrized distance $d_s$ in Eq.\ (\ref{avdistanceerror}) and then deriving a cophenetic distance $d_c$ via the average linkage procedure (\ref{average}) outlined in Sec.\ \ref{sec:Linkage}. Calculation of cophenetic coefficient (\ref{cophenet}) yields the value 
\begin{equation}
c \simeq 0.82, 
\label{cc}
\end{equation}
close to one, so that the ``distances" $d_s$ in Eq.\ (\ref{avdistanceerror}) are close to an ultrametric (\ref{ultram}).

In order to unearth a taxonomic classification of the languages in the set one has to cut the dendrogram at a given level (ultrametric or cophenetic distance) $d_c$. Such a level can be reasonably chosen by relying on a stability criterion of the clustered solution. We therefore  search for a stable partition among the hierarchy yielded by the clustering algorithm, in correspondence to an approximately constant value of the cluster entropy $S$ in a certain range of $d_c$ \cite{kaneko}
\begin{equation}
    S(d_c) = -\sum_{k=1}^{N_{d_c}}P_{d_c}(k)\ln P_{d_c}(k) ~, 
    \label{Boltzmann-entropy}
\end{equation}
where $P_{d_c}(k)$ is the fraction of elements belonging to cluster $k$, $N_{d_c}$ the number of clusters at level $d_c$ in the dendrogram and $0 \leq S(d_c) \leq \ln 21 \simeq 3.04$.

\begin{figure*}
\centering
\includegraphics[width=11cm]{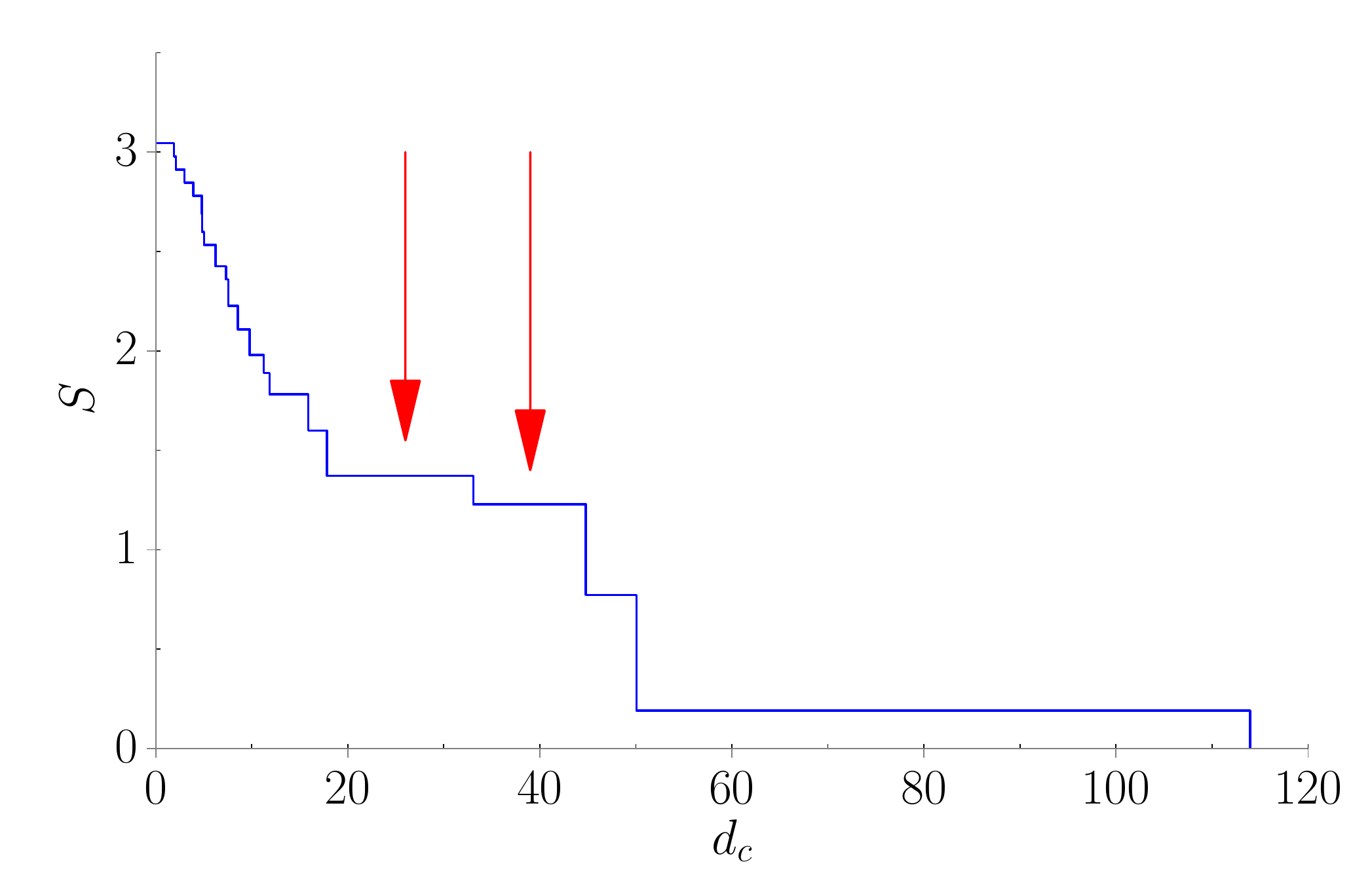}\\
\caption{Cluster entropy across the language tree. The plateaus discussed in the text are indicated by red arrows.}
\label{entropy_average}
\end{figure*}

The entropy (\ref{Boltzmann-entropy}) is plotted in Fig.\ \ref{entropy_average} as a function of $d_c$. The presence of two large adjacent plateaus is manifest. The first one is between $d_c=17.82$ and $d_c=33.08$. If one cuts the dendrogram here, one obtains three large clusters, made up of the Romance (or Latin), Germanic and Slavic languages, and two isolated languages, Greek and Maltese, the latter being a Semitic language (written in Latin script), descended from an Arabic dialect spoken in Sicily and Malta about one thousand years ago. English words make up 15-20\% of the Maltese vocabulary.

The second plateau is between $d_c=33.08$ and $d_c=44.79$. If one cuts the dendrogram here, one obtains three large clusters, made up of the Romance, Germanic (including Maltese) and Slavic languages, and one isolated language, Greek.

The resilience of Greek to join any cluster (see Fig.\ \ref{linkage_alg}) could be ascribed to the fact that we are clustering written languages, in Latin script. For instance, both $\eta$ and $\epsilon$ are translated into the ASCII character for ``e", and both $\omega$ and ``o" (omicron) into the ASCII character for ``o". Presumably, this has an influence on the classification. This influence seems to prevail over the differences of Maltese within the group.
Notice also the very small ``size" of the corresponding point in Fig.\ \ref{table1}, indicating a small error in discriminating Greek within the given language set.

\begin{figure*}
\centering
\includegraphics[width=12 cm]{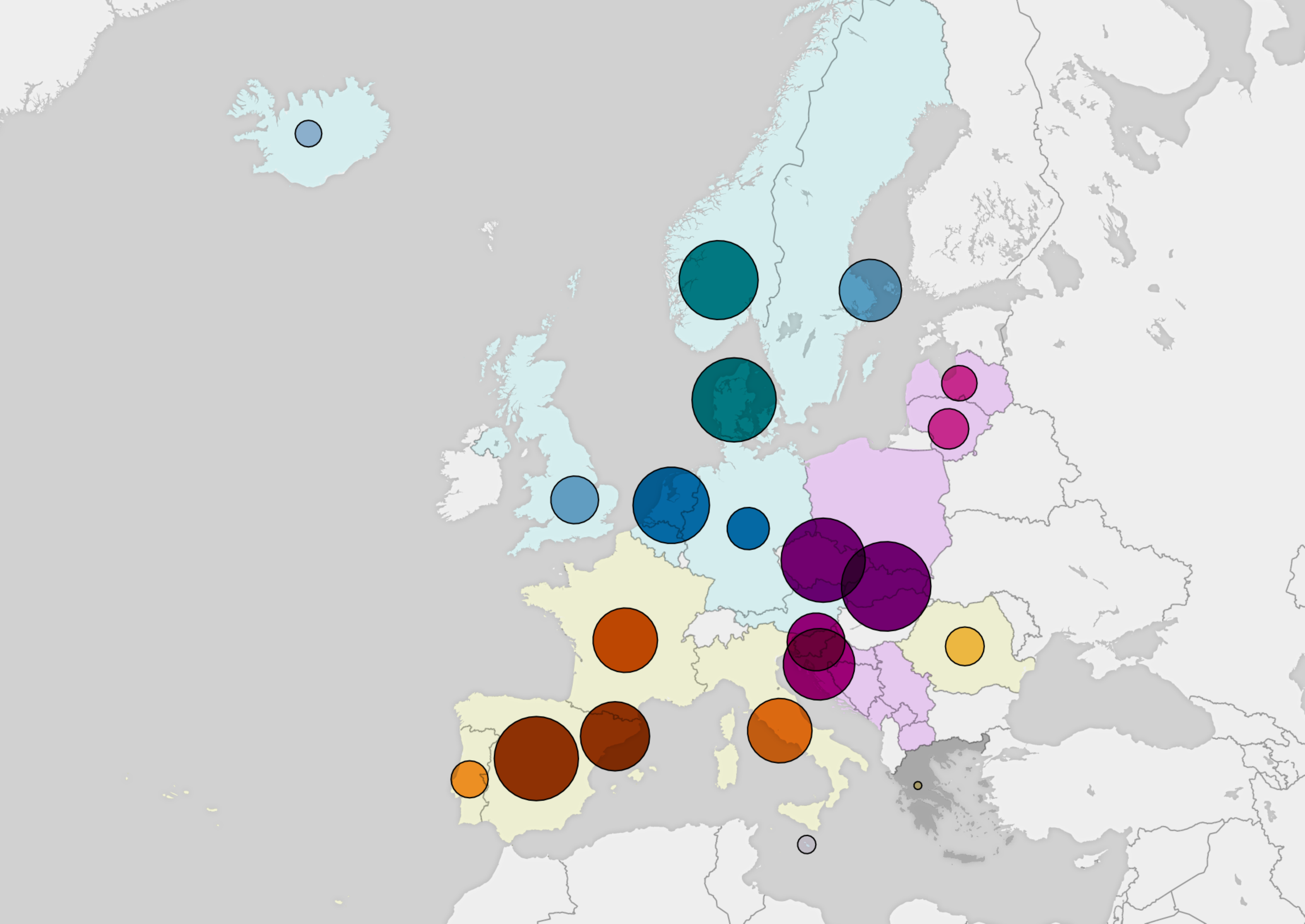}\\
\caption{An artistic view of the linguistic classification. The circles refer to the languages in the data set of Fig.\ \ref{acronyms} (their size being discussed in Sec.\ \ref{sec:dist_lang}). The shades include countries where the language is similar or spoken by part of the population.
(Powered by Mapbox \cite{mapbox}).}
\label{geomap}
\end{figure*}

In Fig.\ \ref{geomap} we summarize our results by making use of a geographical representation. The size of the points is given in the diagonal of the lower panel of Fig.\ \ref{table1}, and is interpreted as the ``overlap" among the given language and the other languages in the set, according to the comments at the end of Sec.\ \ref{sec:dist_lang}.

\section{Conclusions}
\label{sec:concl}

We used a neural network approach to classify a group of 21 Indo-European languages from written texts. We defined a distance among languages, clustered them according to the average linkage procedure and obtained a dendrogram. An ultrametric structure emerged, with a cophenetic coefficient close to unity. By cutting the dendrogram according to an entropic criterion, five subgroups of languages were identiÞed: Romance (or Latin), Germanic and Slavic languages, and two isolated languages, Greek and Maltese. Maltese clusters with the Germanic subgroup if the cut is moved, always in accord with the entropic criterion.

The method we proposed is less efficient for larger language sets. This can be ascribed to the arbitrariness of the definition of distance and the consequent difficulty in satisfying the axioms of distance (to a reasonable level of approximation). No a priori knowledge of the setÕs structure was given or used, in order to carefully avoid the introduction of unwanted bias. Possibly, the performance of the method could be improved by adding extra information.

Our approach might be useful in the fields of data mining and big data analysis and in particular natural language processing. Although we used a simple supervised learning setup, we managed to extract sensible language features from written text. Among future applications one might consider political and biophysical applications, DNA sequence and disease classification and the investigation of the structure of large complex corpora when no or little \textit{a priori} knowledge of the structure is given or available.
Another future challenge would consist in addressing the classification of spoken languages.

\begin{acknowledgments}
Computing resources and technical support were provided by CRESCO/ENEAGRID High Performance Computing infrastructure and staff \cite{cresco1}. CRESCO (Computational RESearch center on COmplex systems) is funded by ENEA and by Italian and European research programs \cite{cresco2}. This work was partially supported by the PRIN Grant No. 2010LLKJBX on ``Collective quantum phenomena: from strongly correlated systems to quantum simulators.". We thank E.\ Marinari and A.\ Monaco for interesting conversations.

\end{acknowledgments}

\appendix

\section{Learning and code}
\label{sec:code}
The code is written in SciPy \cite{scipy} with the MPI4Py package for parallel execution \cite{mpi}. 
It makes use of Theano library \cite{theano,theano2}, originally developed in the field of deep machine learning \cite{deeplearning} and optimized for complex tasks.

\begin{figure*}
\centering
\includegraphics[width=8 cm]{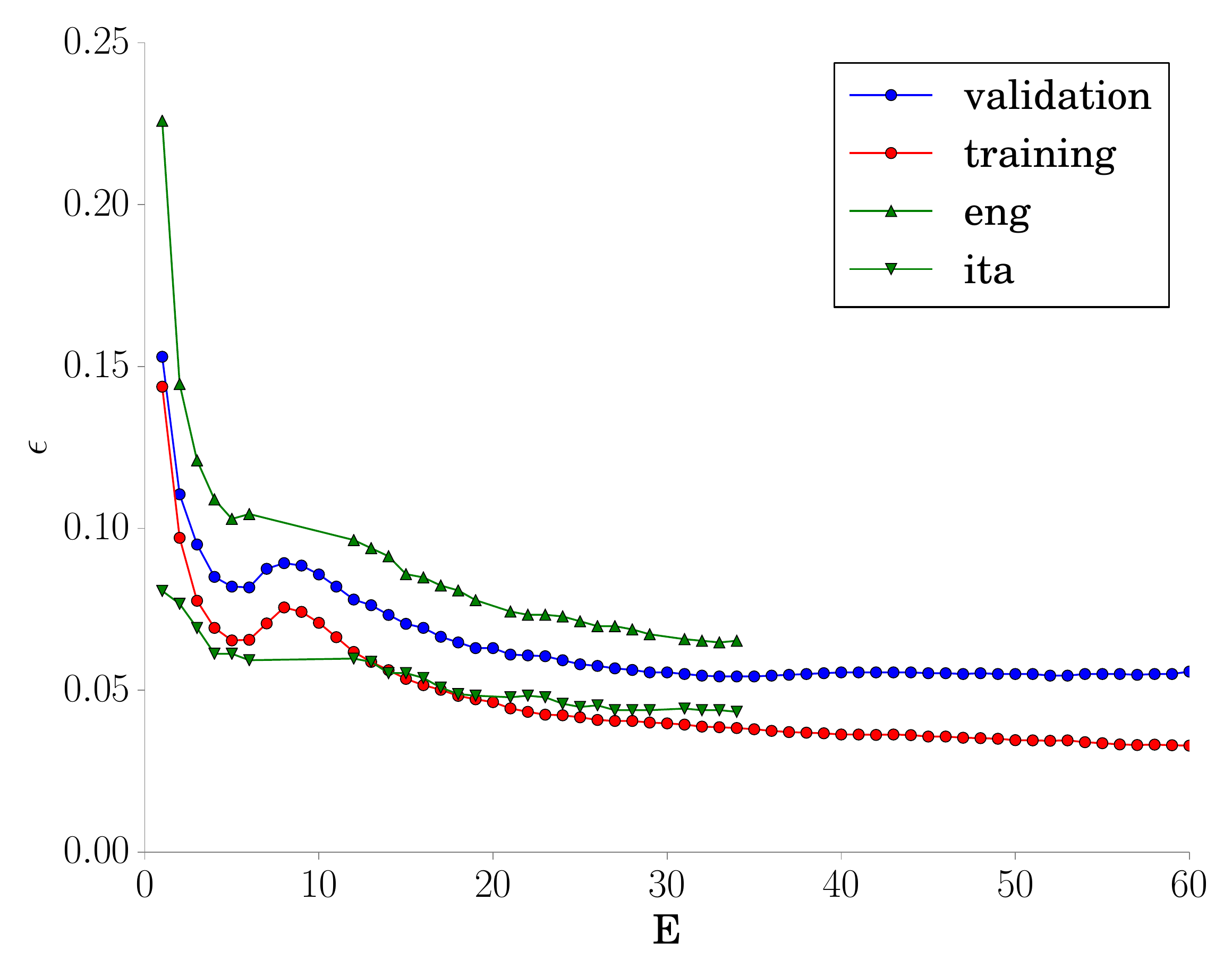} 
\includegraphics[width=8 cm]{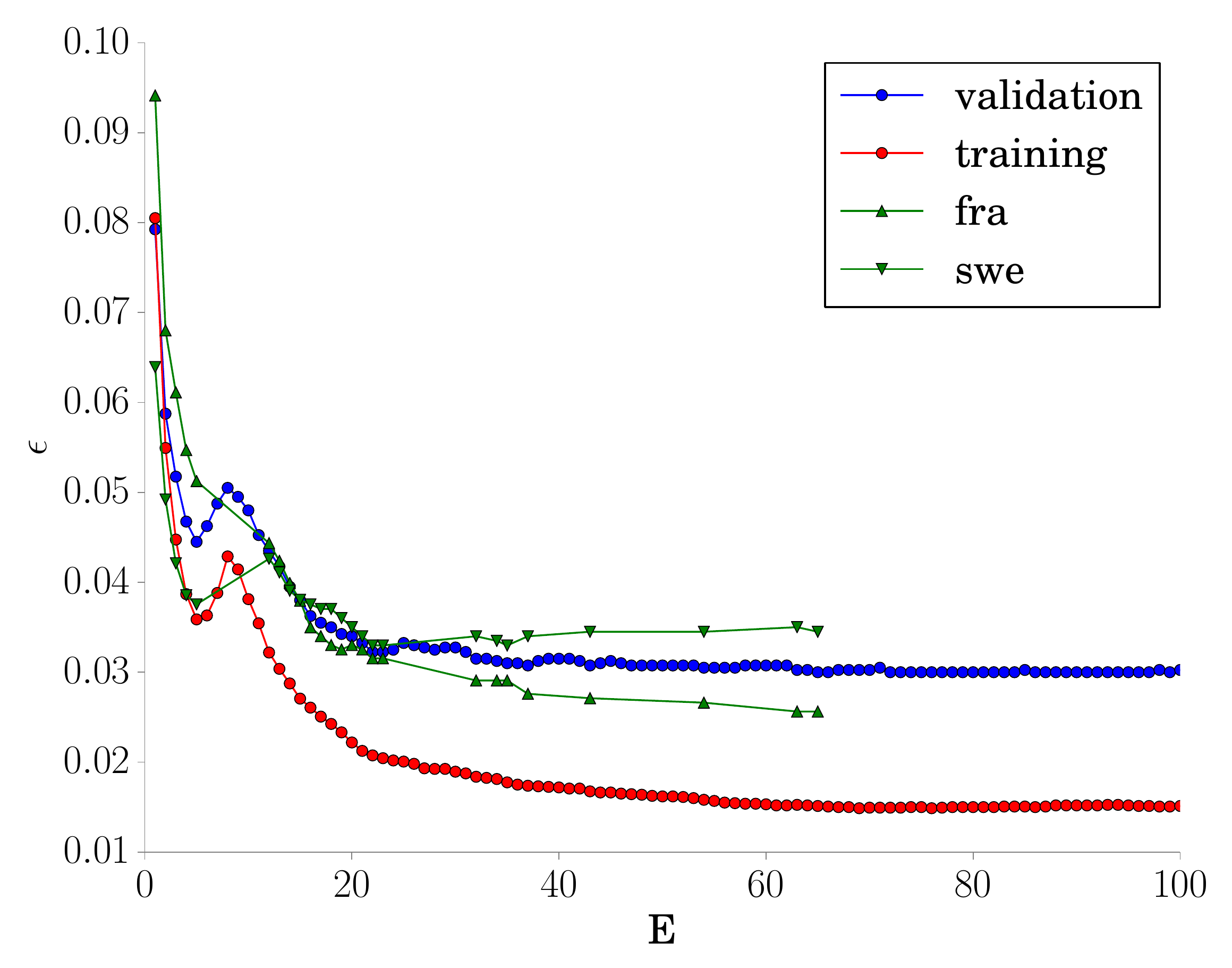}
\caption{Neural network learning. Error vs epoch of training. Upper panel: English vs Italian. Lower panel: French vs Swedish.
Notice that the error on English (namely the percentage of English sentences that are mistakenly classified as Italian) is higher than the error on Italian. By contrast, the error of French vs Swedish is almost symmetric.}\label{epoch}
\end{figure*}

The model implemented is a single-hidden-layer multi-layer perceptron (or artificial neural network ANN), mathematically represented as a function $f: \mathbb{R}^N \to \mathbb{R}^2$
\beq
f(x)=G(b^{(2)} + W^{(2)}(s(b^{(1)} + W^{(1)}x))),
\label{MLPexpr}
\eeq
where $b^{(1)}$ and $b^{(2)}$ are bias vectors, $W^{(1)}$ and $W^{(2)}$ weight matrices, and $G$ and $s$ activation functions. In our case the activation function $s$ is tanh and $G$ is a softmax function. The training process is a supervised learning with back-propagation \cite{backprop} (learning rate $l_r = 0.01$) based on stochastic gradient descent on mini-batches of 20 instances, repeated for almost $E=200$ epochs. The cost function is a negative log likelihood function with two regularization hyper-parameters $l_1=0.001$ and $l_2=0.0001$, related to the $L_1$ norm and squared $L_2$ norm of the weight matrices $W^{(1)}$ and $W^{(2)}$, that avoid an uncontrolled increase of the weights. At every epoch, for every mini-batch, the cost function is evaluated and the ANN parameters are corrected according to the gradient of the cost function, with the learning rate $l_r$ as a multiplying factor. At the end of an epoch, the error between prediction and effective result is computed against the validation set: if it is smaller, it is recorded. Before the $100^{th}$ epoch, the error $\epsilon$ typically becomes steady and the results on the validation set are considered reliable. See Figure \ref{epoch}. The entire process is repeated through a 5-fold cross-validation technique, also by interchanging languages A and B as depicted in Fig.\ \ref{alice}. 

\bibliographystyle{cj}

\end{document}